\documentstyle[12pt,epsf,graphicx]{article}
\newcommand{\bfr}{{\bf r}}

\begin{document}

\title{{\bf Weak point disorder in strongly fluctuating flux-line liquids}}
\author{Panayotis Benetatos$^1$ and M. Cristina Marchetti$^2$ \\
\\
$^1$ Hahn-Meitner-Institut, Abteilung Theoretische Physik (SF5),\\
Glienicker Str. 100, D-14109, Berlin, Germany\\
$^2$ Physics Department, Syracuse University,\\
Syracuse, NY, 13244, USA}
\date{8 April 2005}
\maketitle

\begin{abstract}
We consider the effect of weak uncorrelated quenched disorder
(point defects) on a strongly fluctuating flux-line liquid. We use
a hydrodynamic model which is based on mapping the flux-line
system onto a quantum liquid of {\it relativistic} charged bosons
in $2+1$ dimensions [P. Benetatos and M. C. Marchetti, Phys. Rev.
B {\bf 64}, 054518, (2001)]. In this model, flux lines are allowed
to be arbitrarily curved and can even form closed loops. Point
defects can be scalar or polar. In the latter case, the direction
of their dipole moments can be random or correlated. Within the
Gaussian approximation of our hydrodynamic model, we calculate
disorder-induced corrections to the correlation functions of the
flux-line fields and the  elastic moduli of the flux-line liquid.
We find that scalar disorder enhances loop nucleation, and polar
(magnetic) defects decrease the tilt modulus.



\end{abstract}



\section{Introduction}

The physics of vortex-line arrays in type-II superconductors has
been the subject of intense research activity in the past fifteen
odd years \cite{Blatter,Brandt,Natt,GBhatt,CrabNel}. Potential technological
applications of high-temperature superconductors rely on the
pinning of the vortices in order to eliminate dissipative losses
from their motion. In addition, the competition between  thermal
fluctuations, intervortex interactions, and various types of
disorder gives rise to a variety of phases, thus making vortex
matter an excellent laboratory for the study of fundamental
statistical mechanics.

The effect of disorder  on the Abrikosov lattice phase of vortex
arrays has been the subject of extensive theoretical and
experimental studies \cite{LO,Nat2,FFH,GL,Gold,TKlein}. Much less
understood is the role of disorder in the vortex liquid phase,
which is  known to occupy a broad region of the
magnetic-field-temperature phase diagram in high-temperature
superconductors.  A powerful method for describing both clean and
disordered vortex liquids is the mapping of the statistical
mechanics of directed line liquids in three dimensions onto that
of two-dimensional nonrelativistic quantum bosons
\cite{DRN,MPF,NS}. This mapping assumes small fluctuations of the
flux  lines away from the direction $\hat{\bf z}$ of the external
applied field and intervortex interactions only among line
segments at the same ``height'', $z$. The vortex lines map onto to
the world lines of the quantum particles and this mapping is
particularly useful for describing the effect of columnar pins
caused by heavy-ion irradiation, which correspond to quenched
impurities for the bosons \cite{NV}. In contrast, point defects in
the superconductor correspond to time-dependent disorder for the
bosons and are more difficult to treat in the quantum system. It
has been shown, however, that this type of disorder
 yields ``Lorentzian-squared''
corrections to the thermal static structure function of the vortex
liquid \cite{NLD,TN} and does not renormalize the tilt modulus in
the thermodynamic limit \cite{TN}.

A more convenient framework for describing point disorder is the
hydrodynamic description of directed flux-line liquids developed
some time ago by M. C. Marchetti and D. R. Nelson
\cite{MCMDRNhd,MCMhd}. This earlier implementation of vortex
liquid hydrodynamics takes into account the repulsion among
flux-line segments at different heights, $z$, which renders the
intervortex interaction nonlocal, but is restricted to directed
lines, excluding overhangs or spontaneous vortex loops.

It was argued recently that  vortex loops may be important in the
vortex liquid phase, especially at low fields, where diverging
loop fluctuations may destroy the coherence of the vortex state
via a so-called ``loop blow-out'' transition
\cite{Tes1,Tes2,NSud}. Although earlier numerical evidence for
such a transition in the uniformly frustrated $XY$ model
\cite{NSud} was subsequently shown  to be associated with boundary
effects \cite{OlTeit}, these simulations do indicate that
spontaneous vortex loop formation is important in vortex liquids.
Also, recent experiments in YBCO show evidence of a
liquid-to-liquid transition which appears to agree with the ``loop
blow-out'' scenario \cite{Bouquet}. Disorder, which is ubiquitous
in real high-temperature materials, was not included in the above
mentioned simulations. This renders a detailed comparison with the
experiment problematic.

In this paper we present a hydrodynamic model of strongly
fluctuating flux liquids that allows us to describe in a unified
manner both field-induced  lines and spontaneous loops. The key
ingredient is a new scalar field that describes the density of
flux-line length regardless of its direction or orientation and
therefore incorporates the contribution from neighboring
antiparallel segments associated with closed loos. We use this
framework to study the effect of point disorder on the flux-line
liquid by evaluating perturbatively disorder-induced corrections
to the static correlation functions and to the elastic constants.
Both ``scalar'' point disorder (that couples to flux lines
irrespective of their orientation) and ``vector'' point disorder
(arising from impurities carrying magnetic moments) can naturally
be introduced in the model and play qualitatively different roles
in renormalizing the elastic properties of the liquid. In
particular, ``scalar'' disorder softens the single-line energy,
enhancing the formation of vortex loops.

The statistical mechanics of strongly fluctuating vortex liquids
in zero external magnetic field has been described theoretically
by M. Kiometzis {\it et al.} \cite{Kiom} and by Z. Te{\v
s}anovi{\' c} \cite{Tes1,Tes2} by mapping the vortex line system
onto a collection of \emph{relativistic} two-dimensional quantum
bosons. In this mapping spontaneous vortex loop formation
corresponds to particle-antiparticle creation and annihilation.
Building on this work and its generalization to vortex liquids in
a  field, we recently proposed a general mapping of flux-line
liquids of directed field-induced lines and spontaneous flux loops
onto a two-dimensional system of \emph{relativistic charged}
quantum bosons, where the external applied field for the bosons
corresponds to the bosonic chemical potential \cite{rel}. The main
difference between our model and earlier ones
\cite{Tes1,Tes2,NSud} is that, while these authors consider
fluctuating vortices in a frozen magnetic field (corresponding to
the limit of infinite penetration depth), we describe magnetic
flux lines fluctuating in concert with their vortex cores by
keeping the London penetration depth finite. By coarse-graining
the boson model, we obtained a generalized hydrodynamic
description of strongly fluctuating liquids of flux lines and
loops.

After describing the model in Section 2, we use a Gaussian
approximation of our generalized hydrodynamic free energy to
evaluate the renormalization of the liquid properties from three
different types of uncorrelated disorder in Sections 3, 4 and 5.
Finally, we summarize in Section 6.

\section{The model}

In the hydrodynamic description of a {\it directed} flux-line
liquid proposed by Marchetti and Nelson \cite{MCMDRNhd,MCMhd}, the
large scale properties of the system are described in terms of the
three components of a coarse-grained  vector field, ${\bf t}({\bf
r})=\{{\bf t}_\perp,t_z\}$. The component $t_z$ is the areal
density of flux lines along the direction $z$ of the external
field piercing the $xy$ plane and directed along $\hat{\bf z}$. It
is proportional to the $z$ component of the local magnetic field
in the material. The two-dimensional vector field ${\bf t}_\perp$
describes the local tilt away from the $z$ direction and is
proportional to the local magnetic field in the $xy$ plane. The
correspondence between this hydrodynamic model and the familiar
nonrelativistic boson model is well understood. It is important to
note that the energy of a collection of flux lines in an applied
field has full three-dimensional rotational symmetry. This
symmetry is reduced in the nonrelativistic boson model which is
invariant under Galilean boosts, corresponding to a uniform tilt
of the flux lines away from the direction of the applied field. In
the conventional hydrodynamic free energy, the single-line part is
invariant under uniform tilt (Galilean symmetry), while the
intervortex interaction has full rotational invariance in three
dimensions. Full rotational invariance of the flux-line system
corresponds to Lorentz invariance of the bosons, indicating that
the quantum particles should be treated relativistically.

In Ref. \cite{rel}, we showed that a three-dimensional liquid of
directed field-induced lines and spontaneous flux loops can be
mapped onto a $(2+1)$-dimensional liquid of {\it relativistic
charged} bosons interacting via a screened electromagnetic field.
We exploited this mapping to obtain a general hydrodynamic
description of strongly fluctuating vortex liquids that allows for
overhangs, loop formation, and nonlocal (in $z$) interaction among
vortex segments at different "heights". The hydrodynamic
description of strongly fluctuating vortex liquids is in terms of
{\it four} coarse-grained fields: the three components of the
vector field, ${\bf t}({\bf r})=\{{\bf t}_\perp,t_z\}$, which
measures the local magnetic field in the material, and an
additional scalar field, $\rho({\bf r})$.  The new scalar field
$\rho({\bf r})$ appears because of the possibility of spontaneous
loop formation. It measures the total contour length of flux lines
enclosed in a unit volume centered at point $\bfr$, regardless of
their direction or orientation. Upon coarse-graining,
configurations of antiparallel neighboring flux lines, as arising
from the spontaneous formation of a vortex loop, do not contribute
to ${\bf t}({\bf r})$, but do contribute to $\rho({\bf r})$. In
the relativistic boson mapping, ${\bf t}({\bf r})=\{{\bf
t}_\perp({\bf r}),t_z({\bf r})\}$ corresponds to the {\it
conserved} current-{\it charge}-density, which is a $(2+1)$-vector
with the time-like component $t_z$ the conventional charge density
of bosons and ${\bf t_\perp}$ the charge-current density. In
contrast to the charge density, $t_z$, the number density of
bosons is not conserved due to creation-annihilation events
(corresponding to vortex loop formation in the flux-line system).
The scalar field $\rho({\bf r})$ corresponds to the square of the
amplitude of the complex boson field, $\Phi({\bf r})\equiv
\sqrt{\rho({\bf r})}\exp[i\theta({\bf r})]$. The correlator
$\langle\Phi^*({\bf r}')\Phi({\bf r})\rangle$ is proportional to
the probability of finding a flux line segment connecting the two
points, ${\bf r}$ and ${\bf r}'$ \cite{Kleinert}, irrespective of
the segment's orientation.

The hydrodynamic free energy functional of a strongly fluctuating
flux-line liquid with disorder can be written as ${\cal F}={\cal
F}_0+{\cal F}_D$, where ${\cal F}_0$ is the free energy of the
pure system and ${\cal F}_D$ incorporates the coupling to point
defects. The pure contribution for a line liquid in an external
field, ${\bf H}=H_0\hat{\bf z}$, is  \cite{rel}
\begin{eqnarray}
\label{Fcl} {\cal F}_0[{\bf t}, \rho]&=&
  \int_{\bf r}\bigg\{
      \frac{\epsilon_1}{2}\rho({\bf r})
     +\frac{(k_BT)^2}{8\epsilon_1\rho({\bf r})}[{\bf \nabla}\rho({\bf r})]^2
     +\frac{\epsilon_1}{2 \rho({\bf r})}|{\bf t({\bf r})}|^2
      - \frac{\phi_0}{4\pi}{\bf H}\cdot {\bf t}({\bf r})\bigg\} \nonumber\\
& &   + \frac{1}{2\Omega L}\sum_{\bf q}V(q)|{\bf t}({\bf q})|^2\;,
\end{eqnarray}
where $\epsilon_1=\epsilon_0\ln\kappa$ is the bare flux-line
tension, $\epsilon_0={\phi_0^2}/({4\pi \tilde \lambda})^2$,
$\tilde \lambda$ is the effective London penetration depth, and
$\kappa \equiv \tilde \lambda/\xi$ is the Ginzburg-Landau
parameter, with $\xi$ the coherence length. The last term (written
in Fourier space) describes the intervortex interaction, with
$V(q)=4\pi\epsilon_0\tilde{\lambda}^2/(1+q^2\tilde{\lambda}^2)$.
The average equilibrium values of the hydrodynamic fields are:
${\bf t}_{\perp 0}=0$, $t_{z0}=\pm\rho_0$, and
$\rho_0=(|H_0|-H_{c1})/\phi_0$ for $|H_0|>H_{c1}$, while
$\rho_0=0$ for $|H_0|\leq H_{c1}$. Here
$H_{c1}=4\pi\epsilon_1/\phi_0$ is the lower critical field and
$\rho_0=1/a_0^2$ represents the average areal density of flux
lines in the $xy$ plane, with $a_0$ the intervortex spacing.
Averages over the conformations of flux lines and flux loops are
evaluated with the Boltzmann weight $\sim\exp[-{\cal F}/k_B T]$
{\it and} the constraint
\begin{equation}
\label{constraint} {\bf \nabla}\cdot {\bf t}=0\;,
\end{equation}
which prohibits flux lines from starting or ending within the
sample.
In the relativistic boson mapping it corresponds to charge
conservation. In this paper we present the hydrodynamics of vortex
liquids in isotropic superconductors, but the generalization to
the anisotropic case is straightforward \cite{rel}.

In contrast with the hydrodynamics of {\it directed} flux-line
liquids, the effective free energy of Eq. (\ref{Fcl}) respects the
rotational invariance of a pure flux-line system in an external
field. This rotational invariance corresponds to the Lorentz
invariance of the relativistic boson liquid.  The non-Gaussian hydrodynamic
free energy of {\it directed} vortex liquids \cite{PhyC} is
obtained from Eq. (\ref{Fcl}) simply by setting $\rho({\bf
r})\equiv t_z({\bf r})$. This suggests that the difference
\begin{equation}
\label{tzL} t_z^L({\bf r})\equiv t_z({\bf r})-\rho({\bf r})
\end{equation}
may be interpreted as a density of lines associated with
spontaneous vortex loop fluctuations.

\section{Point defects with magnetic moments parallel to the external field}

Uncorrelated quenched disorder from point material defects is
usually modeled by a random field which couples to the
fluctuations in $t_z({\bf r})$  \cite{NLD}. The disorder
contribution to the free energy is written as
\begin{equation}
\label{dis1} {\cal F}_D=\int d^2r_{\perp}dz V_D({\bf r})\delta
t_z({\bf r})\;.
\end{equation}
The quenched random potential is taken to be Gaussian,
statistically homogeneous and isotropic, with $\overline{V_D({\bf
r})}=0$ and $\overline{V_D({\bf r})V_D({\bf r}')}=\Delta
\delta({\bf r}-{\bf r}')$, where the overbar denotes the quenched
average over disorder. A random field that couples to $t_z$
corresponds to  quenched magnetic dipoles of random strength and
position, with magnetic moments aligned with the $z$ direction. It
favors local alignment of the flux lines with the direction of the
external field (when $V_D({\bf r})<0$).  It corresponds to a
random chemical potential for the nonrelativistic bosons
\cite{TN}, and to a random scalar potential coupling to the charge
density for relativistic charged bosons \cite{rel}.

In order to calculate the disorder-induced contributions to the
various correlation functions characterizing the vortex liquid, we
expand the flux-liquid free energy of Eq.~(\ref{Fcl}) to quadratic
order in the fluctuations of the fields from their equilibrium
values, $\delta {\bf t}({\bf r})={\bf t}({\bf r})-{\bf t}_0$ and $\delta \rho({\bf r}) = \rho({\bf r}) -\rho_0$.
The resulting Gaussian approximation is given by
\begin{eqnarray}
\label{F_Gaussian} {\cal F}^G[\delta{\bf t}, \delta\rho]&\approx&
F'+
     \frac{1}{2\Omega L}\sum_{\bf q}\bigg\{
     \Big[\frac{\epsilon_1}{\rho_0}+\frac{(k_BT)^2}{4\epsilon_1\rho_0}q^2\Big]
                         |\delta\rho({\bf q})|^2
     +\Big[\frac{\epsilon_1}{\rho_0}+V(q)\Big]|\delta{\bf t}({\bf q})|^2
        \nonumber\\
& &   -\frac{\epsilon_1}{\rho_0}\big[\delta t_z({\bf q})\delta\rho(-{\bf q})
                                +\delta t_z(-{\bf q})\delta\rho({\bf q})\big]
        \bigg\}+ {\cal F}_D  \;,
\end{eqnarray}
where $F'$ is the equilibrium value. The Gaussian approximation to
our hydrodynamic model for the pure system can be viewed as a
modified Ornstein-Zernike theory for flux lines \cite{OZ}. The
familiar Gaussian free energy of directed lines is obtained by
setting $\delta\rho({\bf q})\equiv\delta t_z({\bf q})$ in
Eq.~(\ref{F_Gaussian}) and neglecting terms of higher order in
$q$, with the result
\begin{eqnarray}
\label{F_Gaussian_0} {\cal F}^G_d[\delta{\bf t}]\approx F'+
     \frac{1}{2\Omega L\rho_0^2}\sum_{\bf q}\Big\{
     c_{44}^0(q)|\delta{\bf t}_\perp({\bf q})|^2
     +c_{11}^0(q)|\delta t_z({\bf q})|^2
\Big\}+ {\cal F}_D  \;,
\end{eqnarray}
where
\begin{eqnarray}
\label{c440} && c_{44}^0(q)=\rho_0\epsilon_1+\rho_0^2V(q)\;,\\
\label{c110} && c_{11}^0(q)=\rho_0^2V(q)\;, \end{eqnarray}
are the bare local tilt and compressional moduli, respectively, of a
pure {\it directed} flux-line liquid.

The correlation functions of the fluctuations of the hydrodynamic
fields are immediately evaluated by carrying out both a thermal
average with weight $\sim \exp[-{\cal F}^G/k_BT]$ and an average
over quenched disorder. Retaining here and below only the
first-order corrections in $\Delta$, we obtain
%
\begin{eqnarray} \label{tztz} \overline{\langle \delta t_z({\bf
q}) \delta t_z(-{\bf q})\rangle_G}=
   k_BT\frac{\rho_0q_\perp^2}{D({\bf q})}+\Delta\bigg[\frac{\rho_0q_\perp^2}{D({\bf
   q})}\bigg]^2\;,
\end{eqnarray}
\begin{eqnarray}
\label{tperptperp} \overline{\langle t_{\perp i} ({\bf q})
t_{\perp j} (-{\bf q}) \rangle_{G}} &=& k_BT\Big[\frac{\rho_0^2k_B
T}{c_{44}^0(q)}P_{ij}^T({\bf q}_{\perp}) +
\frac{\rho_0q_z^2}{D({\bf q})}P_{ij}^L({\bf
q}_{\perp})\Big]\nonumber\\
&& +\Delta  \Big[\frac{\rho_0 q_\perp q_z}{D({\bf q})}\Big]^2
P_{ij}^L({\bf q}_{\perp}) \;,
\end{eqnarray}
\begin{eqnarray} \label{deltarhodeltarho} \overline{\langle
\delta\rho({\bf q}) \delta\rho(-{\bf q})\rangle_G}&=&
     k_BT\frac{\rho_0q^2}{D({\bf q})}\;
            \frac{c_{44}^0(q)}{\rho_0\epsilon_1(1+q^2a_0^2(a_0/2l_z)^2)}\nonumber\\
     &&       +\Delta \bigg[\frac{\rho_0q_\perp^2}{D({\bf
     q})(1+q^2a_0^2(a_0/2l_z)^2)}\bigg]^2\;,
\end{eqnarray}
where $i, j =(x,y)$, $P_{ij}^L({\bf q}_{\perp})=q_{\perp
i}q_{\perp j}/q_{\perp}^2$ and $P_{ij}^T({\bf
q}_{\perp})=\delta_{ij}-P_{ij}^L({\bf q}_{\perp})$ are
longitudinal and transverse projection operators and
\begin{equation}
D({\bf q})\equiv q_z^2c_{44}^0(q)/\rho_0
  +q_\perp^2\big[c_{11}^0(q)+\delta c_{11}(q)\big]/\rho_0\;.
  \end{equation}
As seen below, $\delta c_{11}(q)$ represents  a short scale
correction to the compressional modulus. It is given by
\begin{equation}
\delta
c_{11}(q)=\rho_0\epsilon_1\frac{q^2a_0^2(a_0/2l_z)^2}{1+q^2a_0^2(a_0/2l_z)^2}\;,
\end{equation}
with $l_z=\epsilon_1a_0^2/(2k_BT)$. The autocorrelation function
of the longitudinal part of ${\bf t}_\perp$ is simply related to
that of $t_z$ as the ``continuity'' equation (\ref{constraint})
requires ${\bf q_\perp}\cdot{\bf t_\perp}=-q_zt_z$. Cross
correlations among the fields are also not vanishing but will not
be given here.

Compressional and tilt deformations of the line liquid can be
described  by defining effective local (wavector-dependent) elastic
moduli as
\begin{equation}
\label{c11q} \frac{\rho_0^2 k_B
T}{{c}_{11}(q_\perp)}=\lim_{q_z \rightarrow 0}
\overline{\langle \delta t_z ({\bf q}) \delta t_z (-{\bf q}) \rangle}\;,
\end{equation}
\begin{equation}
\label{c44q} \frac{\rho_0^2 k_B
T}{{c}_{44}(q_z)}=\lim_{q_\perp \rightarrow 0} P_{ij}^T({\bf
q}_{\perp})\overline{\langle t_i ({\bf q}) t_j (-{\bf q})
\rangle}\;.
\end{equation}
In the absence of disorder, this gives
${c}_{11}(q_\perp)=c_{11}^0(q_\perp)+\delta c_{11}(q_\perp)$,
while ${c}_{44}(q_\perp)=c_{44}^0(q_\perp)$ is unchanged. The
length scale $l_z$ is the thermally-induced entanglement
correlation length introduced by Nelson \cite{DRN}. A single flux
line at finite temperature wanders transversally as it crosses the
sample along the $z$ direction. The entanglement length $l_z$
represents a ``collision'' length in the time-like variable $z$.
It is the length scale a single flux line travels in the $z$
direction before wandering a transverse distance of order $a_0$
and colliding with a neighbor. Such ``collisions'' tend to stiffen
the compressional modulus on small length scales (of order $a_0$)
due to the caging of each line by its neighbors.

The long-wavelength disorder-renormalized elastic constants are defined as
\begin{eqnarray}
\label{c11R} &&\frac{\rho_0^2 k_B T}{c^R_{11}}=\lim_{q_{\perp}
\rightarrow 0} \lim_{q_z \rightarrow 0} \overline{\langle \delta t_z
({\bf q}) \delta t_z (-{\bf q}) \rangle}\;,\\
\label{c44R} &&\frac{\rho_0^2 k_B T}{c_{44}^R}=\lim_{q_z
\rightarrow 0} \lim_{q_{\perp} \rightarrow 0} P_{ij}^T({\bf
q}_{\perp})\overline{\langle t_i ({\bf q}) t_j (-{\bf q})
\rangle}\;,
\end{eqnarray}
where the order of the limits is important. These definitions are
consistent with the definition of the elastic constants as
response functions. Within the Gaussian approximation, point
disorder described as a random field coupled to $t_z({\bf r})$
softens the compressional modulus, but it does not renormalize the
tilt modulus,
\begin{eqnarray}
\label{c11ans}&&
c_{11}^R=\frac{c_{11}^0(0)}{1+\Delta\rho_0^2/(c_{11}^0(0)k_BT)}\;,\\
\label{c44ans}&& c_{44}^R=c_{44}^0(0)\;.
\end{eqnarray}

As discussed in Ref.~\cite{rel}, the long-wavelength limit of the
autocorrelation function of the fluctuations in the field
$t_z^L({\bf r})$, associated with spontaneous vortex loops,
determines the renormalized single-line
energy  $\epsilon_1$, according to
\begin{equation}
\label{tLtLdef} \lim_{q_{\perp}\rightarrow 0}\lim_{q_z\rightarrow
0}\overline{\langle \delta t_z^L({\bf q}) \delta t_z^L(-{\bf
q})\rangle_G}=\frac{\rho_0 k_B T}{\epsilon_1^R} \;,
\end{equation}
irrespective of the order of limits. A random field that couples
to $t_z$  gives no corrections to the line energy and
$\epsilon_1^R=\epsilon_1$. It is important to stress the
difference between the line energy, $\epsilon_1$, and the line
tilt stiffness, $\tilde{\epsilon_1}$. The former is the energy per
unit length needed to create a flux line in the direction of the
external applied field. It is related to the lower
critical field and, in the
boson analogy, it corresponds to the boson rest mass. The tilt
coefficient $\tilde{\epsilon_1}$ measures the stiffness of a line
against a tilt away from the direction of the external field. In
clean, isotropic superconductors, $\epsilon_1=\tilde{\epsilon_1}$
are equal, but they can differ strongly in anisotropic materials
\cite{Kogan}. Using the free
energy of strongly fluctuating flux-line liquids in uniaxial anisotropic
superconductors \cite{rel} (where the external field is applied along the $c$-axis of the material), one can show that the long-wavelength
limit of autocorrelator of the $t_z^L$ field yields $\epsilon_1$,
while the single-vortex part of the tilt modulus yields
$\tilde{\epsilon_1}$. In the following sections, we shall show
that different types of disorder have different effect on these
two distinct quantities.

Finally, the disorder-induced correction to the autocorrelators of
the two different densities, Eqs. (\ref{tztz}) and
(\ref{deltarhodeltarho}), differ at finite wavevectors. This
difference becomes pronounced in strongly entangled liquids (high
temperature and/or density), where $a_0/l_z\sim 1$
($a_0/l_z\approx 0.3$ for $T=77 K$ and $B=2 T$).

\section{Scalar point defects}

In this section we model uncorrelated point disorder as a scalar
random field that couples to the scalar field $\rho({\bf r})$,
\begin{equation}
\label{dis2}
{\cal F}_D=\int d^2r_{\perp}dz V_D({\bf r})\delta \rho({\bf r}) \;.
\end{equation}
This corresponds to a random contribution to the line stiffness
$\epsilon_1$ (or to the quantum particles' mass in the
relativistic charged boson mapping). The random field is assumed
to have zero mean and correlations given by  Eq.~(\ref{disvar}),
although with a different disorder strength, $\Delta$. This random
field physically represents uncorrelated pinning centers which
favor the nucleation of flux lines {\it irrespective} of their
orientation. We therefore expect that it will favor vortex loop
formation. In a liquid of directed lines, there is no difference
between this case and the one discussed in Section 3.

The thermal part of all correlation functions is, of course,
unchanged. Introducing the notation $\overline{\delta \langle ...
\rangle} \equiv \overline{\langle ... \rangle}-\langle ... \rangle$ to
denote the contribution due to disorder to the various correlation
functions, we find
\begin{eqnarray}
\label{1tztz} & &\overline{\delta \langle \delta t_z({\bf q})
\delta t_z(-{\bf q})\rangle_G}=
\Delta\bigg[\frac{\rho_0q_\perp^2}{D({\bf
q})\big({1+q^2a_0^2(a_0/2l_z)^2}\big)}\bigg]^2\;,
\end{eqnarray}
\noindent \begin{eqnarray} \label{1tperptperp} \overline{\delta
\langle t_i ({\bf q}) t_j (-{\bf q}) \rangle_{G}} =\Delta
\bigg[\frac{\rho_0 q_\perp q_z}{D({\bf
q})\big(1+q^2a_0^2(a_0/2l_z)^2\big)}\bigg]^2~P_{ij}^L({\bf
q}_{\perp}) \; ,
\end{eqnarray}
\begin{eqnarray}
\label{1deltarhodeltarho} \overline{\delta \langle \delta\rho({\bf
q}) \delta\rho(-{\bf q})\rangle_G}=\Delta
\bigg[\frac{\rho_0q^2}{D({\bf q})}\;
            \frac{c_{44}^0(q)}{\rho_0\epsilon_1(1+q^2a_0^2(a_0/2l_z)^2)}\bigg]^2
            \;.
\end{eqnarray}

Again, disorder leaves the transverse part of the ${\bf t_\perp}$
correlations unchanged and therefore it does not renormalize the long-wavelength tilt
modulus (in the Gaussian approximation). The renormalized long-wavelength
compressional modulus is identical to that given in
Eq.~(\ref{c11ans}), although the two differ at finite wavelengths.
This type of scalar disorder does, however, yield a finite
renormalization of the line energy that determines  vortex loop
fluctuations. Using the definition given in Eq.(\ref{tLtLdef}), we
find
\begin{equation}
\label{loopcor}
\frac{1}{\epsilon_1^R}=\frac{1}{\epsilon_1}+ \frac{\Delta \rho_0}{k_BT\epsilon_1^2}\;,
\end{equation}
The disorder-induced correction to the line energy on the right
hand side of Eq.~(\ref{loopcor}) can be written as the ratio  of
two length scales as $\epsilon_1^R=\epsilon_1/(1+l_z/l_D)$, where
$l_z$ is the thermal  entanglement length introduced earlier and
$l_D=\epsilon_1^2/(2\Delta \rho_0^2)$ is a characteristic
entanglement length associated with disorder. This type of
quenched disorder softens the line tension, enhancing the
formation of vortex loops. The condition $l_z\approx l_D$ marks a
crossover at $T\sim (\Delta/\epsilon_1)B_z$ from a regime where
loop fluctuations are thermally dominated to one where they are
induced by quenched-disorder. Although numerical studies of the
\emph{pure} uniformly frustrated $XY$ model have ruled out the
scenario of the ``loop blow-out'' transition proposed by Te{\v
s}anovi{\' c} \cite{Tes1,Tes2} and by Sudb{\o} and collaborators
\cite{NSud}, our result indicates that vortex loop formation is
enhanced by disorder, suggesting the possibility that such a
transition may be recovered in sufficiently disordered systems,
where the line tension may be driven to zero.

\section{Point defects with randomly oriented magnetic moments}

Finally, in this section we consider uncorrelated point impurities
that can be modeled as quenched magnetic dipoles with random
orientation. This corresponds to a vector disorder potential that
couples with statistically equal strength to all components of the
local magnetic field, yielding
\begin{equation}
\label{dis3}
{\cal F}_D=\int d^2r_{\perp}dz {\bf V}_D({\bf r})\cdot{\bf t}({\bf r}) \;,
\end{equation}
with $\overline{V_{D \mu}({\bf r})}=0$ and $\overline{V_{D
\mu}({\bf r})V_{D \nu}({\bf r}')}=\delta_{\mu \nu} \Delta
\delta({\bf r}-{\bf r}')$, for $\mu, \nu=(x,y,z)$. The random
field favors alignment of the local magnetic field due to flux
lines with uncorrelated quenched random directions.

To Gaussian order, the disorder-induced corrections to the
correlation functions are
\begin{eqnarray} \label{2tztz} & &\overline{\delta
\langle \delta t_z({\bf q}) \delta t_z(-{\bf q})\rangle_G}=
\Delta\bigg[\frac{\rho_0q_\perp q}{D({\bf q})}\bigg]^2 \;,
\end{eqnarray}
\noindent \begin{eqnarray} \label{2tperptperp} & &\overline{\delta
\langle t_i ({\bf q}) t_j (-{\bf q}) \rangle_{G}} =\Delta
\bigg\{\bigg[\frac{\rho_0^2}{c_{44}^0(q)}\bigg]^2P_{ij}^T({\bf
q}_{\perp}) +\bigg[\frac{\rho_0q_\perp q}{D({\bf q})}\bigg]^2
P_{ij}^L({\bf q}_{\perp})\bigg\} \;,
\end{eqnarray}
\begin{eqnarray}
\label{2deltarhodeltarho} \overline{\delta \langle \delta\rho({\bf
q}) \delta\rho(-{\bf q})\rangle_G}= \Delta
\bigg[\frac{\rho_0q_{\perp}q}{D({\bf
q})\big(1+q^2a_0^2(a_0/2l_z)^2\big)}\bigg]^2\;.
\end{eqnarray}

The disorder-induced correction to the oriented areal density
autocorrelation, Eq (\ref{2tztz}), differs significantly from that
obtained for the other two types of disorder when
$q_{\perp}\approx 0$ and $q_z\neq 0$.

As in the case presented in Section 3, this type of disorder does
not renormalize the line stiffness $\epsilon_1$.  The transverse
part of the tangent field autocorrelator does, however, acquire a
finite correction, which leads to a renormalization of the
long-wavelength tilt modulus. From Eq. (\ref{c44R}), we find
\begin{equation}
\label{dc44} \frac{1}{c_{44}^R}=\frac{1}{c_{44}^0}+\frac{\Delta}{k_B
T} \frac{1}{\big[\epsilon_1 + \rho_0\phi_0^2/(4 \pi)\big]^2}\;.
\end{equation}
As shown by Larkin and Vinokur \cite{LV}, the renormalization of
$c_{44}$ can be related to that of its single-vortex part,
$c_{44}^v=\rho_0\tilde\epsilon_1$, with $\tilde\epsilon_1$ the
tilt stiffness  of a \emph{directed} line. This is renormalized by
the ``vector'' disorder according to
\begin{equation}
\label{epsilon1tilde}
 \tilde\epsilon_1^R=\epsilon_1(1-l_z/l_D)\;.
 \end{equation}
Isotropic random dipoles tend to align  flux-line segments along
the random directions of the magnetic moments carried by the
impurities and facilitate tilting away from the $z$-direction. On
the other hand, the line tension defined in Eq.~(\ref{tLtLdef})
and associated with loop fluctuations remains unrenormalized by
this type of disorder. In both cases, the disorder-induced
corrections are determined by the ratio $l_z/l_D$ of the thermal
and disorder entanglement lengths (see Eq.~(\ref{loopcor})).

Finally, the disorder-induced
renormalization of the compressional modulus is the same as for
the other two types of point disorder in the long-wavelength
limit, although it differs at finite wavevectors.

\section{Conclusions}

In this paper, we have calculated weak-disorder-induced
corrections to the static correlation functions of strongly
fluctuating (arbitrarily curved and looping) flux-line liquids
using a Gaussian hydrodynamic approach. Three realizations of
point disorder have been considered, corresponding to magnetic and
non-magnetic defects.  Our results for the structure function, the
scalar density ($\rho({\bf r})$) autocorrelator, and the tilt
field (${\bf t}_{\perp}({\bf r})$) correlations can, in principle,
be probed experimentally using polarized and unpolarized neutron
scattering. We have predicted qualitative differences between
these quantities at finite wavevectors for different types of
disorder which are experimentally testable. From these
correlations, we have also obtained the disorder-induced
corrections to the elastic moduli. The compressional modulus gets
softened for all three types of point disorder. Loop fluctuations
are enhanced by scalar disorder. The tilt modulus is softened by
isotropic random polar (magnetic) disorder.

A softening of the elastic constants by quenched disorder may
appear surprising, as  naively one would expect pinning to stiffen
the system. A result qualitatively similar to ours has, however,
been obtained before in a different system \cite{RadNel}.
Preliminary results of a calculation of the disorder-induced
correction to the compressional modulus taking into account
non-Gaussian terms in the free energy suggest that this correction
may change sign depending on the system parameters (temperature or
magnetic field), indicating that a perturbative approach may not
adequate in this case \cite{unpub}. If the softening were to
survive a non-perturbative calculation, this would signal a
disorder-driven instability of the vortex liquid phase. Clearly
more work beyond the Gaussian approximation is needed to answer
these questions.

\vspace{0.2in} MCM was supported by the National Science
Foundation under grant DMR-0305407.


\begin{thebibliography}{}

\bibitem{Blatter}
G. Blatter, M. V. Feigel'man, V. B. Geshkenbein, A. I. Larkin, and V. M. Vinokur, Rev. Mod. Phys. {\bf 66} 1125 (1994).


\bibitem{Brandt}
E. H. Brandt, Pep. Prog. Phys. {\bf 58} 1465 (1995).


\bibitem{Natt}
T. Nattermann and S. Scheidl, Adv. Phys. {\bf 49} 607 (2000).

\bibitem{GBhatt}
T. Giamarchi and S. Bhattacharya, in {\it High Magnetic Fields: Applications to condensed matter physics and spectroscopy}, C. Bertier {\it et al.}, eds., p. 314 (Springer-Verlag, 2002).


\bibitem{CrabNel}
G. W. Crabtree and D. R. Nelson, Phys. Today {\bf 50}(4), 38 (1997).

\bibitem{LO}
A. I. Larkin and Y. N. Ovchinikov, J. Low Temp. Phys. {\bf 34}, 409 (1979).


\bibitem{Nat2}
T. Nattermann, Phys. Rev. Lett. {\bf 64}, 2454 (1990).


\bibitem{FFH}
D. S. Fisher, M. P. A. Fisher, and D. A. Huse, Phys. Rev. B {\bf 43}, 130 (1990).



\bibitem{GL}
T. Giamarchi and P. Le Doussal, Phys. Rev. B {\bf 52}, 1242 (1995).


\bibitem{Gold}
Y. Y. Goldschmidt, Phys. Rev. B {\bf 56}, 2800 (1997).


\bibitem{TKlein}
T. Klein {\it et al.}, Nature {\bf 413}, 404 (2001), and references therein.

\bibitem{DRN}
D. R. Nelson, Phys. Rev. Lett. {\bf 60}, 1973 (1988).


\bibitem{MPF}
M. P. A. Fisher and D. H. Lee, Phys. Rev. B {\bf 39}, 2756 (1989).

\bibitem{NS}
D. R. Nelson and H. S. Seung, Phys. Rev. B {\bf 39}, 9153 (1989).


\bibitem{NV}
D. R. Nelson and V. N. Vinokur, Phys. Rev. B {\bf 48}, 13060 (1993).

\bibitem{NLD}
D. R. Nelson and P. Le Doussal, Phys. Rev. B {\bf 42}, 10113 (1990).

\bibitem{TN}
U. C. T{\"a}uber and D. R. Nelson, Phys. Rep. {\bf 289}, 157 (1997).

\bibitem{MCMDRNhd}
M. C. Marchetti and D. R. Nelson, Phys. Rev. B {\bf 42}, 9938 (1990).

\bibitem{MCMhd}
M. C. Marchetti and D. R. Nelson, Physica C {\bf 174}, 40 (1991).

\bibitem{Tes1}
Z. Te{\v s}anovi{\' c}, Phys. Rev. B {\bf 51}, 16304 (1995).

\bibitem{Tes2}
Z. Te{\v s}anovi{\' c}, Phys. Rev. B {\bf 59}, 6449 (1999).

\bibitem{NSud}
A. K. Nguyen and A. Sudb{\o}, Phys. Rev. B {\bf 60}, 15037 (1999).

\bibitem{OlTeit}
P. Olsson and S. Teitel, Phys. Rev. B {\bf 67}, 144514 (2003).

\bibitem{Bouquet}
F. Bouquet {\it et al.}, Nature {\bf 411}, 448 (2001).

\bibitem{Kiom}
 M. Kiometzis, H. Kleinert, and A. M. J. Schakel, Fortschr. Phys. {\bf 43}, 697 (1995).

\bibitem{rel}
P. Benetatos and M. C. Marchetti, Phys. Rev. B {\bf 64}, 054518 (2001).

\bibitem{Kleinert}
H. Kleinert, {\it Gauge Fields in Condensed Matter}, (World Scientific, Singapore, 1989).

\bibitem{PhyC}
P. Benetatos and M. C. Marchetti, Physica C {\bf 332}, 237 (2000).

\bibitem{OZ}
L. S. Ornstein and F. Zernike, Proc. Acad. Sci. Amst. {\bf 17}, 793 (1914).

\bibitem{Kogan}
V. G. Kogan, Phys. Rev. B {\bf 24}, 1572 (1981).


\bibitem{LV}
A. I. Larkin and V. M. Vinokur, Phys. Rev. Lett. {\bf 75}, 4666 (1995).

\bibitem{RadNel}
L. Radzihovsky and D. R. Nelson, Phys. Rev. A {\bf 44}, 3525 (1991).

\bibitem{unpub}
P. Benetatos and M. C. Marchetti (unpublished).


\end{thebibliography}
\end{document}